\documentclass[12pt,english]{article}

\usepackage[T1]{fontenc}
\usepackage[latin9]{inputenc}
\usepackage[a4paper]{geometry}
\usepackage{color}
\definecolor{document_fontcolor}{rgb}{0, 0, 0}
\color{document_fontcolor}
\usepackage{stackrel}
\usepackage{setspace}
\makeatletter
\@ifundefined{date}{}{\date{}}

\@ifundefined{definecolor}
 {\usepackage{color}}{}
\usepackage{array}\usepackage{amsthm}\usepackage{float} \makeatletter

\newcommand{\lyxmathsym}[1]{\ifmmode\begingroup\def\b@ld{bold}
  \text{\ifx\math@version\b@ld\bfseries\fi#1}\endgroup\else#1\fi}

 \@ifundefined{definecolor}{

}{\makeatletter} \AtBeginDocument{
  
}

\usepackage{fancyhdr}\usepackage{titlesec}\titleformat{\chapter}[display]
  {\normalfont\huge\bfseries\centering}{\MakeUppercase\chaptertitlename\ \thechapter}{20pt}{\LARGE}

\usepackage[square, numbers, comma, sort&compress]{natbib}

\makeatother

\usepackage{babel}
\begin{document}

\title{Improvement on Quantum bidirectional teleportation $2\leftrightarrow2$
or $2\leftrightarrow3$ qubit teleportation protocol via 6-qubit entangled
state}

\author{Mitali Sisodia\thanks{\emph{Email: mitalisisodiyadc@gmail.com}}}
\maketitle
\begin{center}
Department of Physics, Indian Institute of Technology Delhi,
\par\end{center}

\begin{center}
New Delhi 110016, India
\par\end{center}
\begin{abstract}
Recently, Ri-Gui Zhou et al. {[}Int. J. Theor. Phys.\textbf{ 59},
166\textendash 172 (2020){]} proposed a scheme for bidirectional quantum
teleportation of two$\leftrightarrow$two and two$\leftrightarrow$three
qubit states by utilizing a six-qubit entangled state as a quantum
channel. It is observed that the intrinsic efficiency calculated by
Zhou et al. is not correct. In this improved scheme, we show that
bidirectional teleportation of two$\leftrightarrow$two and two$\leftrightarrow$three
qubit quantum states can be done by using optimized quantum resource
and less consumption of classical resource. We also calculate the
intrinsic efficiency of our improved scheme which is much better than
the Zhou's scheme. Additionally, we also discuss about the security
of the protocol.
\end{abstract}
Controlled NOT (CNOT). Bell state measurement. 

\section{Introduction}

In 1993, Bennett et al. presented a first theoretical concept of the
transferring of an unknown quantum state with the use of two-qubit
entangled quantum channel and some of classical communication and
the name of the scheme is quantum teleportation (QT) \cite{bennett}.
Afterwards, numerous schemes of QT \cite{QT1,QT2,Ex1,Ex2,ex3,mitali,IBM,light}
and modified QT have been proposed, such as bidirectional teleportation
\cite{gen1,bi,gen2,sisodia,c7,comment}, controlled quantum teleportation
\cite{c1,c2,c3,four,security,MARLON,MEILING,seven_qubit}, bidirectional
controlled teleportation \cite{c4,c5,c8} with the use of multi-qubit
entangled quantum states. In 2017, Mitali et al. \cite{mitali} proposed
the optimized QT scheme of $n$-qubit quantum state of the form of
$|\psi\rangle=\stackrel[i=1]{m}{\sum}\alpha_{i}|x_{i}\rangle$ and
shown that multi-qubit quantum state (specific forms) can be teleported
by using some of Bell states. After that, several work for variants
of QT have been proposed with the use of optimized quantum resources
and less consumption of the classical resource \cite{gen2,sisodia,MEILING}
(and references therein). 

Following the trend, not long ago, a scheme on bidirectional teleportation
of two$\leftrightarrow$two and two$\leftrightarrow$three qubit quantum
state by using six-qubit entangled quantum resource has been proposed
by Zhou et al. \cite{comment}. In Zhou's work \cite{comment}, we
found that unnecessary quantum resource is used and calculated intrinsic
efficiency is not correct. Also, they have not discussed anywhere
about the security of the protocol against attacks. So, we have improved
Zhou's recent scheme \cite{comment},  which is proposed by using
multi-qubit entangled state as a quantum channel. In this improved
version, we have shown that six-qubit multi-qubit entangled state
is not necessary, only two sets of Bell states or four-qubit entangled
state is sufficient for the bidirectional teleportation of two$\leftrightarrow$two
and two$\leftrightarrow$three qubit quantum states. To boost the
quality of this improved work, we have calculated the intrinsic efficiency
of the protocol and discussed about the security of the scheme. 

This paper is organized as follows: In Section-\ref{sec:Improvement-on-Zhou's},
improvement of Zhou's scheme \cite{comment} is discussed. Efficiency
and security analysis of the protocol is discussed in Section- \ref{sec:Intrinsic-efficiency}
and \ref{sec:Security-analysis}and finally, we have summarized the
discussion in the conclusion section. 

\section{Improvement on Zhou's scheme \label{sec:Improvement-on-Zhou's}}

Let us review the Zhou's bidirectional scheme \cite{comment}, a six-qubit
entangled state is used as a quantum channel for the bidirectional
teleportation of two$\leftrightarrow$two and two$\leftrightarrow$three
qubit states. 

The states are,

\begin{equation}
\begin{array}{ccc}
|\psi\rangle_{A} & = & \alpha_{A}|00\rangle_{a_{1}a_{2}}+\beta_{A}|11\rangle_{a_{1}a_{2}},\\
|\psi\rangle_{B} & = & \alpha_{B}|00\rangle_{b_{1}b_{2}}+\beta_{B}|11\rangle_{b_{1}b_{2}},
\end{array}\label{eq:1}
\end{equation}

and

\begin{equation}
\begin{array}{ccc}
|\psi\rangle_{A} & = & \alpha_{A}|00\rangle_{a_{1}a_{2}}+\beta_{A}|11\rangle_{a_{1}a_{2}},\\
|\psi\rangle_{B} & = & \alpha_{A}|000\rangle_{b_{1}b_{2}b_{3}}+\beta_{A}|111\rangle_{b_{1}b_{2}b_{3}},
\end{array}\label{eq:2}
\end{equation}

The quantum channel is,

\begin{equation}
\begin{array}{ccc}
|\phi\rangle_{123456} & = & \frac{1}{2}\left(|000000\rangle+|001011\rangle+|110100\rangle+|111111\rangle\right)_{123456}\end{array}\label{eq:3-1}
\end{equation}

where coefficients $\alpha$ and $\beta$ are satisfying the normalization
condition $|\alpha|^{2}+|\beta|^{2}=1$ and qubits $A$ and $B$ belong
to Alice and Bob. In Zhou's scheme, Alice and Bob performs GHZ-state
measurements on their qubits. 

In this improved version of Zhou's scheme, the used quantum channel
is,

\begin{equation}
|\varphi\rangle_{A_{1}B_{1}A_{2}B_{2}}=\frac{1}{2}\left(|0000\rangle+|0011\rangle+|1100\rangle-|1111\rangle\right)_{A_{1}B_{1}A_{2}B_{2}}\label{eq:4}
\end{equation}

To teleport bidirectionally two and three-qubit quantum state of the
form of Eq. (\ref{eq:1}) and (\ref{eq:2}) with the help of four-qubit
entangled state (see Eq. (\ref{eq:4})), Alice and Bob applies one
and two CNOT operations on their qubits. Whereas $a_{1}$ and $b_{1}$
play a role as control qubits and remaining qubits $a_{2}$, $b_{2},$
$b_{3}$ as target qubits. After applying CNOT operations the states
become a single-qubit quantum states and the rest is registered qubits. 

Now the task is the teleportation of single-qubit quantum state simultaneously
to Alice to Bob, Bob to Alice. In this scheme, Alice and Bob perform
Bell-state measurements on their qubits and some unitary operations
to recover an unknown quantum states. Complete steps of bidirectional
teleportation of single-qubit quantum state with the use of same quantum
channel (see Eq. (\ref{eq:4})) is clearly shown in detail in paper
\cite{gen2}. In the last step, Alice and Bob introduce auxiliary
in state $|0\rangle$to recover the initial quantum states (see Eq.
(\ref{eq:1}) and (\ref{eq:2})). 

\section{Security analysis \label{sec:Security-analysis}}

In this improved scheme, we checked the security of the Zhou's improved
scheme against communication attacks- intercept-and-resend attack
and the entangle-and-measure attack. For security purpose, we introduce
decoy-qubit technology, which secures our protocol against attacks.
In these communication attacks, Eve (enemy) tries to steal the information
in many ways- 

Eve intercepts Alice's and Bob's transmitting qubits and resend fake
qubits to Alice and Bob and she tries to get entangled with the Alice's
and Bob's qubit and performs the measurement. 

To overcome these attacks or to find the Eve's presence, we add decoy
qubits with the transmitting qubits and the point is - Eve does not
know the position of the decoy qubits. Consequently, Alice and Bob
will know the presence of Eve with the help of decoy-qubit technology,
nocloning theorem and measurement method. Therefore, attacks are  ineffective
to the scheme.

\section{Intrinsic efficiency \label{sec:Intrinsic-efficiency}}

In this section, we calculated the intrinsic efficiency (IE) of this
improved scheme of Zhou's scheme by using the formula $\eta=\frac{q_{i}}{q_{r}+c_{r}+a_{u}}$,
where $q_{i}$ denotes the number of qubits consisting of the quantum
information to be transferred, $q_{r}$ represents the number of qubits
used in the quantum channel, $c_{r}$ is the number of bits used in
the classical communication process to complete the task, and $a_{u}$
is the number of auxiliary qubits \cite{seven_qubit}. In Zhou's work
\cite{comment}, they have calculated IE $\sim$40\% and $\sim$45.5\%
for the first and second part (two$\leftrightarrow$two and two$\leftrightarrow$three
qubit teleportation), respectively.  But, we have observed that calculated
IE by Zhou et al. is not correct. In their scheme, $q_{i}$= 4 and
5 (two$\leftrightarrow$two and two$\leftrightarrow$three qubit)
$q_{r}$= 6 (six-qubit used in channel), $c_{r}$= 6 (six classical
bits is used by Alice and Bob (GHZ state measurement is performed
by Alice and Bob)), $a_{u}$= 1 (for second part (two$\leftrightarrow$three
qubit)), therefore IE will be $\sim$33.3\% and $\sim$38\% for the
first and second part, respectively. 

In our improved scheme, we calculate the IE for the first (two$\leftrightarrow$two)
and second part ( two$\leftrightarrow$three) and the value is $\sim$40\%
and $\sim45\%$, which is much better than the Zhou's scheme \cite{comment}.
Consequently, this improved scheme on Zhou's scheme possesses higher
intrinsic efficiency than the Zhou et al.'s work.

\section{Conclusion}

Bidirectional teleportation is a two party scheme in which both parties
send an unknown quantum states to each other simultaneously. Recently,
Zhou et al. \cite{comment} proposed a bidirectional scheme by using
six-qubit entangled quantum state. In this work, we have shown that
six-qubit entangled state is not necessary, four-qubit state is sufficient
for the teleportation of two$\leftrightarrow$two and two$\leftrightarrow$three
qubit states. However, it is found that calculated IE in the Zhou's
scheme is not correct. Therefore, we have calculated the correct IE
for the Zhou's work and found that IE of our improved scheme is much
better than the Zhou's work \cite{comment}. It can be concluded that
the improved scheme possess higher efficiency and low quantum cost
as compare to Zhou's proposed work. This work is not only applicable
to the scheme of Zhao et al., it is applicable to many other schemes
in which unnecessary (higher than the minimum required amount) quantum
resources have been used.

\end{document}